\documentclass[aps,prl,twocolumn,groupedaddress]{revtex4-1}
\usepackage{graphicx}
\usepackage{amsmath}
\usepackage{ulem}
\usepackage{color}

\begin{document}

%\preprint{}

\title{Effect of quenched disorder on a quantum spin liquid state of triangular-lattice antiferromagnet 1T-TaS$_{2}$
	%Elucidating the nature of a quantum spin liquid state of triangular-lattice antiferromagnet 1T-TaS$_{2}$
}


\author{H. Murayama$^{1}$, Y. Sato$^{1}$, T. Taniguchi$^{1}$, R. Kurihara$^{1}$, X. Z. Xing$^{1}$, W. Huang$^{1}$,
\\S. Kasahara$^{1}$, Y. Kasahara$^{1}$, I. Kimchi$^{2}$, M. Yoshida$^{3}$, Y. Iwasa$^{3,4}$, 
Y. Mizukami$^{5}$, T. Shibauchi$^{5}$, 
M. Konczykowski$^{6}$, Y. Matsuda$^{1}$}
%\thanks{}
\affiliation{$^{1}$Department of Physics, Kyoto University, Kyoto 606-8502, Japan\\
$^{2}$JILA, NIST and Department of Physics, University of Colorado, Boulder, USA\\
$^{3}$RIKEN Center for Emergent Matter Science (CEMS), Wako 351-0198, Japan\\
$^{4}$Quantum-Phase Electronics Center and Department of Applied Physics, the University of Tokyo, Tokyo 113-8656, Japan\\
$^{5}$Department of Advanced Materials Science, University of Tokyo, Chiba 277-8561, Japan\\
$^{6}$Laboratoire des Solides irradi\'{e}es, CEA/DRF/lRAMIS, Ecole Polytechnique, CNRS, Institut Polytechnique de Paris, F-91128 Palaiseau, France 
%Laboratoire des solides irradi\'{e}es, L'\'{E}cole polytechnique, France
}


\date{\today}
\begin{widetext}
\begin{abstract}
A quantum spin liquid (QSL) is an exotic state of matter characterized by quantum entanglement and the absence of any broken symmetry.  A long-standing open problem, which is a key for fundamental 
understanding the mysterious QSL states,  is how the %ground state and elementary excitations are affected by 
quantum fluctuations respond to randomness due to 
quenched disorder.  Transition metal dichalcogenide 1T-TaS$_2$  is a candidate material that hosts a QSL ground state with spin-1/2 on the two-dimensional perfect triangular lattice. Here, we performed systematic studies of low-temperature heat capacity and thermal conductivity on pure, Se-substituted and electron irradiated crystals of 1T-TaS$_{2}$, where the substitution of S by Se induces 
weak disorder and electron irradiation induces 
strong quenched disorder.   In pure 1T-TaS$_{2}$, the linear temperature term of the heat capacity %specific heat 
$\gamma T$  and the finite residual linear term of the thermal conductivity in the zero-temperature limit $\kappa_{0}/T\equiv\kappa/T(T\rightarrow0)$ are %both 
clearly resolved,  consistent with the presence of gapless spinons with a Fermi surface.  Moreover, while the strong magnetic field slightly enhances $\kappa_0/T$, it strongly suppresses $\gamma$.  These unusual contrasting responses to magnetic field imply the coexistence of two types of gapless excitations with itinerant and localized characters.  Introduction of additional weak random exchange disorder in 1T-Ta(S$_{1-x}$Se$_x$)$_2$ leads to vanishing of $\kappa_0/T$, indicating that the itinerant gapless excitations are sensitive to the disorder.  On the other hand, in both pure and Se-substituted systems,  the magnetic contribution of the heat capacity obeys a universal scaling relation, which is consistent with a theory that assumes the presence of localized orphan spins forming random singlets.   These results appear to capture an essential feature of the QSL state of 1T-TaS$_2$;  localized orphan spins induced by disorder form random valence bonds and are surrounded by a QSL phase with spinon Fermi surface.   Electron irradiation in pure 1T-TaS$_2$  largely enhances $\gamma$ and changes the scaling function dramatically, suggesting a possible new state of spin liquid. 

\end{abstract}
\end{widetext}

%\keywords{}

\maketitle

\section{I. INTRODUCTION}

A spin liquid is a state of matter in which the spins are correlated, yet fluctuate strongly like in a liquid down to very low temperatures. 
In particular, a quantum spin liquid (QSL) is a novel state in which enhanced quantum fluctuations prevent the system from the long-range magnetic ordering even at zero temperature \cite{spinon_Balents,Zhou2017}, forming a complex superposition of singlet states.  Since these states never order, they have no broken symmetries and %thus 
are not described by the Landau's theory of phase transitions.  The ground states of  the QSLs are quantum-mechanically entangled and are expected to host fractional quasiparticle excitations. 
The notion of the QSLs is firmly established in one-dimensional (1D) spin systems as well as in their ladder cousins.  In 2D and 3D systems, on the other hand,  
it is widely believed that QSLs are usually realized in the presence of competing orders or geometrical frustrations. %
Despite tremendous efforts during the past several decades, however, the true nature of the ground states of QSL candidate materials has still remained elusive. One of the most important and long-standing open problems, which is a key for understanding the mysterious QSL states both theoretically and experimentally, is how the quantum fluctuations respond to local randomness induced by quenched disorder, such as defects/impurities.

Of specific interest in 2D frustrated spin systems has been the quantum spin-1/2 triangular-lattice Heisenberg antiferromagnet, which has been suggested to  be a very
prototype of a QSL in resonating valence bond (RVB) model with a quantum superposition of spin singlets \cite{RVB, RVB2}. Unfortunately, only a few candidates of the QSL states have been reported for 2D triangular lattice, including organic Mott insulators, $\kappa$-(BEDT-TTF)$_{2}$Cu$_{2}$(CN)$_{3}$ \cite{BEDT-TTF_NMR,BEDT-TTF_TC} (hereafter abbreviated as BEDT-TTF), EtMe$_{3}$Sb[Pd(dmit)$_{2}$]$_{2}$ \cite{dmit_NMR1, dmit_NMR2} (DMIT) and $\kappa$-H$_{3}$(Cat-EDT-TTF)$_{2}$ \cite{Hcat,HCat2,HCat3} (H-Cat) and inorganic YbMgGaO$_{4}$ \cite{Yb_C, Yb_crystal,Yb_spinon}. In particular, in the above organic systems, no magnetic ordering occurs at least down to 1/1000 of $J/k_{B}=$ 200 - 300 K ($J$ is the exchange interaction between neighboring spins) \cite{BEDT-TTF_NMR, dmit_NMR1}. 

The most remarkable and intriguing feature in the QSL state on triangular lattice is that in DMIT and H-Cat the heat capacity has a non-zero linear %-in-
temperature term $\gamma T$ and thermal conductivity has a finite residual linear term $\kappa_{0}/T\equiv\kappa/T(T\rightarrow0)$ \cite{dmit_k, dmit_C,HCat2,HCat3}.   (Recent measurements report that there are two types of DMIT with and  without finite $\kappa_{0}/T$ \cite{Taillefer_2019_DMIT, Li_2019_DMIT, Yamashita_2019_DMIT}.)  This demonstrates the emergence of gapless spin excitations that behave like mobile charge carriers in a paramagnetic metal with a Fermi surface, although the charge degrees of freedom are gapped.  The gapless excitations has also been reported by magnetic susceptibility measurements \cite{Watanabe_torque}. The observed highly mobile and gapless excitations have been discussed in terms of fermionic spinons, fractionalized particles that carry spin but no charge, that form the Fermi surface \cite{Motrunich2005,Lee2005}.   On the other hand, the finite residual linear term $\kappa_{0}/T$ is not observed in thermal conductivity of BEDT-TTF and YbMgGaO$_{4}$\cite{,BEDT-TTF_TC, Yb_K}, even though the non-zero linear %temperature 
term $\gamma T$ is observed in heat capacity \cite{BEDT-TTF_C, Yb_C}.   

The effect of quenched disorder has been a fundamental and longstanding issue of the QSL ground states, which has been extensively addressed theoretically.   Randomness had been considered to be detrimental  for QSL states because it can localize the resonating singlet spins.   In recent years, on the other hand, it has been pointed out that strong randomness can act as an additional frustration, which stabilizes QSL states \cite{Watanabe2014}.  Moreover, it has been suggested that even weaker bond randomness can affect quantum paramagnetic phases such as QSLs by inducing orphan spins, as emergent spin defects created by the surrounding  valence bonds \cite{VBS_defect}.  

However, despite its importance,  the systematic effect of randomness on the QSL ground states has been little explored experimentally because of the lack of appropriate systems.   In the organic compounds, for example, the triangular lattice consisting of large molecular dimers with spin-1/2 is distorted from the perfect triangular lattice. Moreover, the glassy dielectric behavior observed in the organic salts suggests the charge imbalance within large molecular dimers, whose influence on the QSL state is unclear.  In YbMgGaO$_{4}$, Mg$^{2+}$/Ga$^{3+}$ site mixing introduces strong randomness in the exchange interaction between spin-1/2 on Yb sites. Moreover, in YbMgGaO$_{4}$ with $J/k_{B}\sim4$\,K,   low energy excitations are difficult to extract at low enough temperature, $T\ll J/k_{B}$ \cite{Yb_C, Yb_crystal,Yb_spinon}. The presence of randomness is also pointed out in QSL candidates on other model systems 
such as herbertsmithite ZnCu$_{3}$(OH)$_{6}$Cl$_{2}$,  which consists of Cu kagom\'{e} layers separated by nonmagnetic Zn layers.  In this system, random substitution of magnetic Cu$^{2+}$ by nonmagnetic Zn$^{2+}$ (and vice versa) together with a possible Jahn-Teller distortion induces the effective quenched randomness \cite{Kagome}.	Furthermore, in the above systems,  degree of randomness cannot be controlled.  Thus the situation calls for a new system.

\begin{figure}[t]
	\includegraphics[width=\columnwidth]{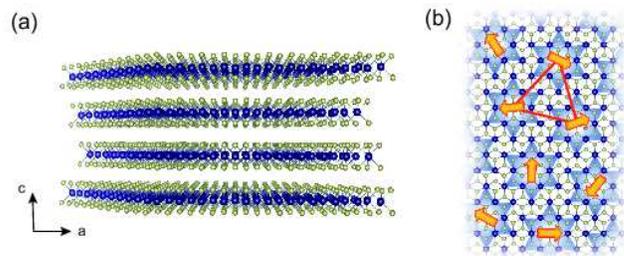}
	\caption{(a) Crystal structure of  1T-TaS$_2$. Each layer consists of tantalum atoms (purple) sandwiched by sulfur atoms (yellow). (b) A schematics of the star-of-David clusters, which appear as a result of the C-CDW transition.  In the Mott insulating state,  the electrons localized at the centers of the clusters form $S = 1/2$ 2D perfect triangular lattice.}
\end{figure}

Recently, transition metal dichalcogenide 1T-TaS$_{2}$ has aroused great interest as a candidate material that hosts a QSL ground state on the 2D perfect triangular lattice \cite{TaS2_Law}. This compound 
is a layered material that contains one Ta layer sandwiched by two S layers; these layers are weakly coupled by van der Waals interactions, as illustrated in Fig.1(a). The Ta atoms form a 2D triangular lattice. At high temperatures ($T>550$ K), 1T-TaS$_{2}$ is metallic. As the temperature is lowered, it exhibits 
an incommensurate CDW phase below 550 K, followed by a nearly commensurate CDW (NC-CDW) phase below 350 K. It undergoes a commensurate CDW (C-CDW) transition at 180 K, below which the unit cell is reconstructed into a rotated triangular lattice characterized by $\sqrt{13}\times\sqrt{13}$ structure described as star-of-David clusters with 13 Ta atoms \cite{TaS2_CDW, TaS2_C_chi}. Strong electron correlation gives rise to a Mott insulating state, in which one localized electron resides at the center of the star-of-David cluster. As a result, a perfect 2D triangular lattice with $S=1/2$ is formed at low temperatures, as illustrated in Fig.\,1(b).  The nuclear magnetic resonance (NMR) experiments have reported the power law dependence of spin-lattice relaxation rate $1/T_1$, indicating that the system is not in the band insulating state in which $1/T_1$ decays exponentially.   Muon spin relaxation and nuclear quadrupole resonance (NQR) experiments have reported no long-range magnetic ordering down to 20\,mK, despite 
%\sout{the strong exchange coupling $J/k_{B}>1000$ K} 
%\textcolor{red}{
the strong exchange coupling $J/k_{B}$ which is at least hundreds of Kelvin as suggested by magnetic susceptibility measurements 
%} 
%\textcolor{red}{the fact that the exchange coupling $J/k_{B}$ is at least hundreds of Kelvin as suggested by magnetic susceptibility measurements} 
\cite{TaS2_Law,TaS2_muon_NQR}.   
Moreover, gapless behavior of the spin dynamics has also been suggested by the muon and NQR measurements. 

\begin{figure}[t]
\includegraphics[width=\columnwidth]{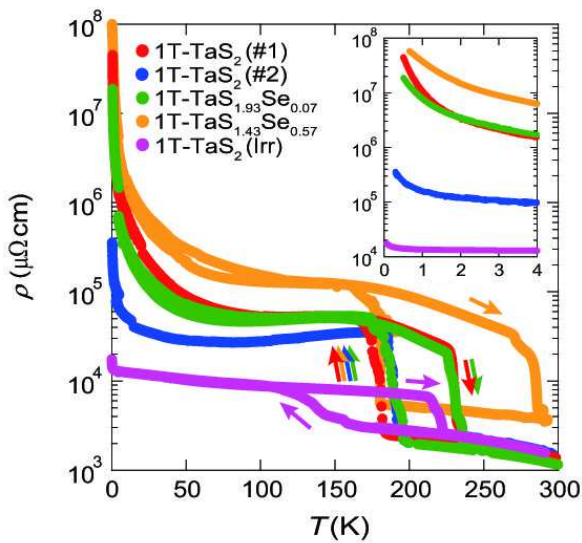}
\caption{Temperature dependences of the resistivity in single crystals of pure 1T-TaS$_{2}$\#1 (red), 1T-TaS$_{2}$\#2 (blue), 1T-TaS$_{1.93}$Se$_{0.07}$ (green), 1T-TaS$_{1.43}$Se$_{0.57}$ (orange) and electron-irradiated 1T-TaS$_{2}$(Irr) (purple). The resistivity is measured for both cooling and heating processes except for 1T-TaS$_{2}$(\#2). For 1T-TaS$_{2}$(\#2), data obtained for the cooling process is shown. The inset shows the resistivity below 4 K.}
\end{figure}

To reveal the nature of QSL states, it is crucially important to clarify whether the low-lying excitations are gapped or gapless, and whether they are localized or itinerant. The heat capacity %specific heat 
$C$ and thermal conductivity $\kappa$ measurements provide crucial information on these issues. The former includes both localized and itinerant excitations, while the latter sensitively probes elementary itinerant excitations and is totally insensitive to localized ones such as those responsible for Schottky contributions, which contaminates the heat capacity measurements at low temperatures.  The $T$-linear term in the heat capacity $\gamma T$ has been reported, indicating the presence of gapless excitations \cite{TaS2_muon_C}.  The measurements of the thermal conductivity, on the other hand,  has reported no discernible residual linear term in zero field, $\kappa_{0}/T=\kappa/T(T\rightarrow0)=0$, indicating the absence of itinerant magnetic excitations \cite{TaS2_k}.    Based on these results, the presence of the spinons that form a Fermi surface has been questioned. 

 In this paper, to obtain new insights into low-lying spin excitations of 1T-TaS$_2$,  the effect of randomness induced by quenched disorder on the spin liquid state is investigated.  For this purpose,  we performed  systematic studies of low-temperature heat capacity and thermal conductivity on pure, Se-substituted and electron irradiated crystals of 1T-TaS$_{2}$.  Substitution of S by Se in the neighboring S-layers induces the random exchange disorder in magnetic Ta-layers, which would act as a weak disorder on the QSL. On the other hand, electron irradiation introduce atomic-scale point defects into Ta layers, which act as strong quenched disorder in 1T-TaS$_2$.  
In fact, as shown later, a small number of vacancies at Ta site (2 vacancies per 10$^5$ Ta atoms) dramatically influence the properties of QSL state.  
We report that the itinerant gapless excitations coexist with the localized gapless excitations.  We find that the former excitations are strongly suppressed by the disorder, while the latter excitations are robust against the disorder.   Moreover, the heat capacity %specific heat 
arising from the latter excitations obeys a universal scaling relation, which assumes the presence of orphan spins forming random long range valence bonds.  Based on these results, we argue that localized orphan spins form random singlet valence bonds and are surrounded by a QSL phase with spinon Fermi surface.  Moreover,  introduction of strong disorder by electron irradiation leads to a possible new state of a spin liquid.

\section{II. SAMPLE PREPARATION AND METHODS}
High quality 1T-TaS$_2$, 1T-TaS$_{2-x}$Se$_{x}$ ($x$=0.07 and 0.57)  single crystals were grown by the chemical vapor transport method.  To introduce spatially homogeneous strong disorder, we employed 2.5-MeV electron beam irradiation at the SIRIUS Pelletron linear accelerator operated by the Laboratorie des Solides Irradi\'ees (LSI) at Ecole Polytechnique. This incident energy is sufficient to form vacancy-interstitial (Frenkel) pairs, which act as point-like defects. 
%To prevent the point-defect clustering, we performed irradiation at 25\,K using a $H_2$ recondenser. 
With displacement energy of 10\,eV, irradiation of 20\,mC/cm$^2$ dose causes $\sim2$ vacancies per 10$^5$ Ta sites, which is about three times larger than the number of vacancies for S sites. 

Figure\,2 depicts the temperature dependence of the resistivity $\rho(T)$ for crystals of 1T-TaS$_{2}$(\#1 and \#2), 1T-TaS$_{1.93}$Se$_{0.07}$, 1T-TaS$_{1.43}$Se$_{0.57}$  and electron-irradiated 1T-TaS$_{2}$(Irr). 
For 1T-TaS$_{2}$(\#1), 1T-TaS$_{1.93}$Se$_{0.07}$, 1T-TaS$_{1.43}$Se$_{0.57}$, and 1T-TaS$_{2}$(Irr), $\rho(T)$ was measured both on cooling and on warming.  The hysteresis around 200\,K is due to the first order phase transition between NC-CDW and C-CDW \cite{TaS2_CDW, TaS2_C_chi}. 
\textcolor{black}{The width of the hysteresis loop of the resistivity is enhanced in 1T-TaS$_{1.43}$Se$_{0.57}$ and electron-irradiated 1T-TaS$_2$, indicating that the first-order phase transition is broadened by the disorder.  All the crystals show the insulating behavior at low temperatures, but there is no clear trends in the amplitude of the resistivity on the defect levels. It has been suggested that stacking faults of TaS$_2$ layers or remaining domain boundaries of NC-CDW % domain boundaries between C-CDW and remaining NC-CDW 
give rise to the filamentary metallic conduction at low temperatures \cite{Zwick98}, although their contributions to the heat capacity is negligibly small. Therefore, the magnitude of the resistivity at low temperature may not be directly related to the defect level. We also note that in TaS$_2$ where the finite residual $T$-linear thermal conductivity observed, the Lorentz number $L =(\kappa_0/T)\rho$ estimated from $\kappa_0/T \sim 0.05$\,W/K$^2$m and $\rho=10^6-10^8$\,$\mu\Omega$cm at the lowest temperature is $\sim 10^4-10^6$  times larger than that of a conventional metal $L_0=\pi^2/3(k_B/e)^2=2.44\times10^{-8}$\,W$\Omega$/K$^2$, indicating a spectacular violation of Wiedemann-Franz law. This shows that the filamentary metallic conduction is not relevant to the finite residual $T$-linear thermal conductivity. } 
% The fact that the irradiated crystal shows the largest change in the temperature dependence of resistivity from the pristine sample indicates that the point defects in the Ta site introduced by electron irradiation act as much stronger disorder than the S-site chemical substitution.

\begin{figure*}[t]
\includegraphics[width=2\columnwidth]{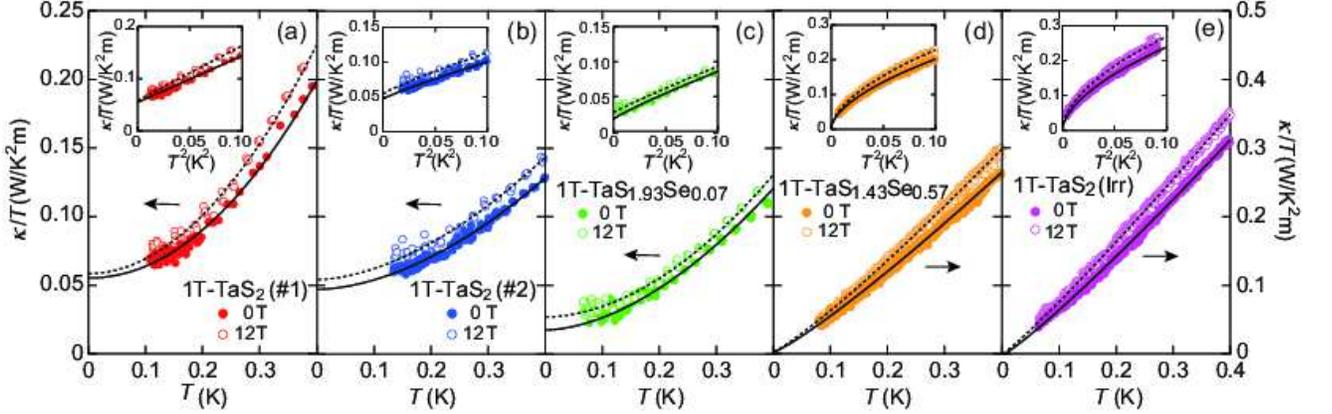}
\caption{Temperature dependences of $\kappa/T$ for (a) 1T-TaS$_2$\#1, (b) 1T-TaS$_2$\#2, (c) 1T-TaS$_{1.93}$Se$_{0.07}$, (d) 1T-TaS$_{1.43}$Se$_{0.57}$, and (e) 1T-TaS$_2$(Irr) in zero magnetic field (filled circles) and at $\mu_{0}H=$ 12\,T applied along $c$ axis (open circles). The insets show $\kappa/T$ plotted as a function of $T^{2}$. Solid and dotted lines in (a)-(c) show the fits by the formula $\kappa/T = \kappa_{0}/T + b_1T^{2}$ for 0\,T and 12\,T, respectively.  Solid and dotted lines in (d) and (f) show the fits by  $\kappa/T = \kappa_{0}/T + b_2T^{p}$ with $p\approx$1 for 0\,T and 12\,T, respectively.  The residual value $\kappa_{0}/T$($\approx 0.05$\,W/K$^2$m) for 1T-TaS$_2$ \#1 and \#2 is largely  suppressed in 1T-TaS$_{1.93}$Se$_{0.07}$ ($\kappa_0/T\approx 0.02$\,W/K$^2$m). In 1T-TaS$_{1.43}$Se$_{0.57}$ and 1T-TaS$_2$(Irr)  $\kappa_0/T$ is absent. }
\end{figure*}

The thermal conductivity was measured down to 80\,mK by the standard one-heater-two-thermometers steady-state technique.  The crystal sizes of 1T-TaS$_2$\#1, 1T-TaS$_2$\#2,  1T-TaS$_{1.93}$Se$_{0.07}$, 1T-TaS$_{1.43}$Se$_{0.57}$, and 1T-TaS$_{2}$(Irr) are 
$0.42$ (distance between contacts) $\times 0.26$ (width) $\times 0.025$ (thickness),  $0.98\,\times 0.250\,\times 0.042, 1.7\times  0.37 \times 0.080, 0.81\times 0.26\times0.073$ and $ 0.45\times 0.76 \times 0.023$\, mm$^3$, respectively.  
%$0.416\times 0.261\times 0.025,  0.975\times 0.250\times 0.042, 1.649\times  0.367 \times 0.080, 0.810\times 0.257\times0.073$ and $ 0.447\times 0.756 \times 0.023$\, mm$^3$, respectively.  
The heat current was applied parallel to the 2D plane and the magnetic field was applied perpendicular to the 2D-plane.  As shown in Fig.\,2, the magnitude of the resistivity at low temperature is sample dependent.  Therefore,  it is crucially important to measure the thermal conductivity and the heat capacity on the  same crystal.     To measure the heat capacity of tiny crystals used for the thermal conductivity experiments,   we used a long relaxation method \cite{Kasahara2018}.   For both measurements, the crystals were cooled down very slowly, as the first-order CDW transition is sensitive to the cooling rate and transition is smeared out by a rapid cooling.

 %[1?5]. The heat conductivity isformed primarily by both acoustic phonons and itinerantspinons, while the latter form QSL. Since the phonon contribution is insensitive to the applied magnetic field B, the elementary excitations of QSL can be further explored by the magnetic-field dependence of k.

\section{III. RESULTS AND DISCUSSION}
\subsection{Thermal conductivity of 1T-TaS$_{2}$}

%#1(0T) : $\kappa_{0}/T=0.055, \kappa_{ph}/T\sim T^{2.03}$
%#1(12T): $\kappa_{0}/T=0.058, \kappa_{ph}/T\sim T^{1.96}$
%#2(0T) : $\kappa_{0}/T=0.047, \kappa_{ph}/T\sim T^{1.78}$
%#2(12T): $\kappa_{0}/T=0.054, \kappa_{ph}/T\sim T^{1.74}$
%Se0.07(0T) : $\kappa_{0}/T=0.026, \kappa_{ph}/T\sim T^{1.90}$
%Se0.07(12T): $\kappa_{0}/T=0.027, \kappa_{ph}/T\sim T^{1.88}$
%Se0.57(0T) : $\kappa_{0}/T=0.002, \kappa_{ph}/T\sim T^{1.15}$
%Se0.57(12T): $\kappa_{0}/T=0.003, \kappa_{ph}/T\sim T^{1.15}$
%Irr(0T) : $\kappa_{0}/T=-0.003, \kappa_{ph}/T\sim T^{1.12}$
%Irr(12T): $\kappa_{0}/T=-0.001, \kappa_{ph}/T\sim T^{1.09}$

Figures\,3(a)-(e) show the  thermal conductivity divided by temperature $\kappa/T$ plotted as a function of temperature for 1T-TaS$_2$\#1, \#2,  1T-TaS$_{1.93}$Se$_{0.07}$, 1T-TaS$_{1.43}$Se$_{0.57}$, and 1T-TaS$_{2}$(Irr), respectively,  at low temperatures in zero field (filled circles) and at $\mu_0H=$12\,T (open circles) applied perpendicular to the 2D plane.   The insets show the same data plotted as a function of $T^{2}$.   As the temperature is lowered, $\kappa/T$ decreases in proportion to $T^2$.   On the other hand, as shown in the main panels and insets of Figs.\,3(d) and (e), $\kappa/T$ for 1T-TaS$_{1.43}$Se$_{0.57}$ and 1T-TaS$_2$(Irr) decreases as $\kappa/T\propto T^{p}$ with $p\approx 1$.  The best fit is obtained by $p=1.15$ and $1.09-1.12$ for  1T-TaS$_{1.43}$Se$_{0.57}$ and 1T-TaS$_2$(Irr), respectively.     In insulating magnets,  thermal conductivity can be written as a sum of the spin and phonon contributions, $\kappa=\kappa_{ph} +\kappa_{spin}$.  When the mean free path of acoustic phonon $\ell_{ph}$ exceeds the average size of the samples at low temperature, the  phonons undergo specular or diffuse scattering from a crystal boundary.   The phonon conductivity in the boundary scattering regime is expressed as $\kappa_{ph}=\frac{1}{3}C_{ph}\langle v_{s}\rangle \ell_{ph}$, where $C_{ph}=\beta T^3$ is the phonon heat capacity of the Debye model with  coefficient $\beta$ and  $\langle v_{s}\rangle$ is the mean acoustic phonon velocity. For diffuse scattering limit due to rough surface roughness, $\ell_{ph}$ becomes $T$-independent, resulting in $\kappa_{ph}\propto T^{3}$.  For specular reflection limit, on the other hand,  $\ell_{ph}$ depends on $T$ as  $\ell_{ph}\propto 1/T$, leading to $\kappa_{ph}\propto T^{2}$. In real systems, the phonon conductivity depends on $T$ as $\kappa_{ph}\propto T^{\alpha}$ with $\alpha$ of intermediate value between 2 and 3. Thus, 1T-TaS$_2$ (\#1 and \#2) and 1T-TaS$_{1.93}$Se$_{0.07}$ crystals are close to the diffuse scattering limit, while 1T-TaS$_{1.43}$Se$_{0.57}$ and electron irradiated 1T-TaS$_2$ crystals are close to specular reflection limit. 
\textcolor{black}{These appear to be consistent with the fact that  $\kappa_{ph}$ of 1T-TaS$_2$  and 1T-TaS$_{1.93}$Se$_{0.07}$ is smaller than $\kappa_{ph}$ of 1T-TaS$_{1.43}$Se$_{0.57}$ and electron irradiated 1T-TaS$_2$, because $\ell_{ph}$ in the specular reflection limit can be enhanced from $\ell_{ph}$ in the diffuse scattering limit.  } 
 
\begin{figure*}[t]
\includegraphics[width=2\columnwidth]{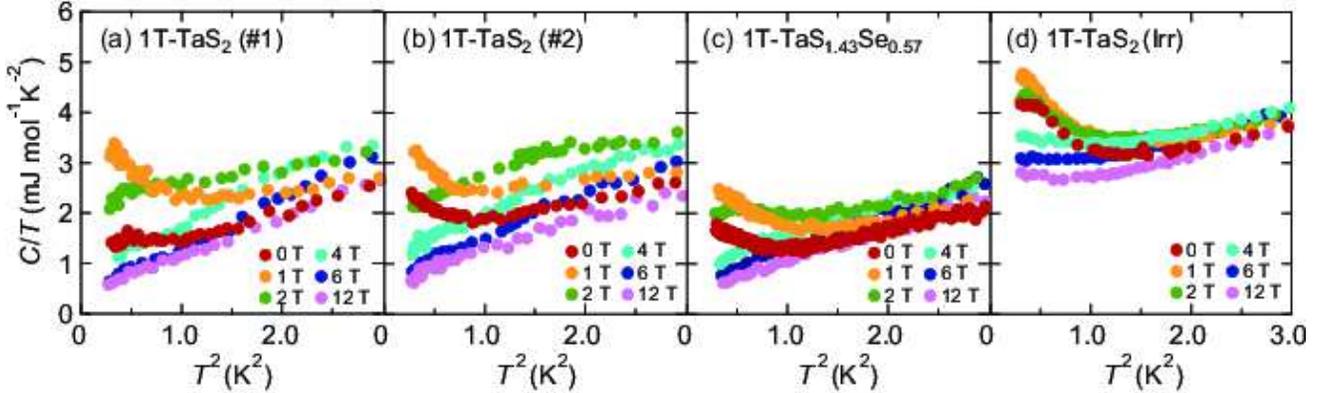}
\caption{$C/T$ plotted as a function of $T^2$ for (a) 1T-TaS$_2$\#1, (b) 1T-TaS$_2$\#2, (c) 1T-TaS$_{1.43}$Se$_{0.57}$, and (d) 1T-TaS$_2$(Irr) in zero field and at $\mu_{0}H=$1, 2, 4, 6 and 12\,T. For all the samples, an upturn of $C/T$ is observed at low temperatures at low fields.}
\end{figure*}

%\begin{figure*}
%\includegraphics[width=2\columnwidth]{Fig5.eps}
%\caption{$(C - C_S)/T$ plotted as a function of $T^2$ for (a) 1T-TaS$_2$\#1, (b) 1T-TaS$_2$\#2, (c) 1T-TaS$_{1.43}$Se$_{0.57}$, and (d) 1T-TaS$_2$(Irr) in zero field and at $\mu_{0}H=$1, 2, 4, 6 and 12\,T. }
%\end{figure*}

The finite intercept,  shown by both plots of $\kappa/T$ vs. $T$ and $\kappa/T$ vs. $T^{2}$ depicted in Figs.\,3(a), (b) and (c) and their insets,  provides evidence for finite $\kappa_0/T$, demonstrating the presence of mobile and gapless spin excitations.  This provides support on  the presence of emergent spinons that form a Fermi surface \cite{Motrunich2005,Lee2005,He_spinon}.  We note that the observed finite  $\kappa_{0}/T$ rules out a possibility of  spin liquids with gapless Dirac-like spinons.    The magnitude of $\kappa_{0}/T$($\approx 0.05$\,W/K$^2$m) for both 1T-TaS$_2$ \#1 and \#2 is largely  suppressed by Se-substitution in 1T-TaS$_{1.93}$Se$_{0.07}$ ($\kappa_0/T\approx 0.02$\,W/K$^2$m).  Moreover, as shown in Fig.\,3(d), both plots of $\kappa/T$ vs. $T$ and $\kappa/T$ vs. $T^{2}$  show vanishing of $\kappa_0/T$  in 1T-TaS$_{1.43}$Se$_{0.57}$.  These results indicate that even weak disorder introduced by Se-substitution in the neighboring layers of magnetic Ta-layers leads to a large suppression of conduction of  the itinerant quasiparticles.    In 1T-TaS$_2$(Irr) the strong disorder from irradiation also gives rise to vanishing $\kappa_0/T$.  In an applied magnetic field, while $\kappa_{0}/T$ is almost unchanged within error bars, $\kappa/T$  is slightly enhanced at finite temperatures, which  may be attributed to the enhancement of phonon mean free path due to the suppression of the spin-phonon scattering by external magnetic field.    

The observed finite  $\kappa_{0}/T$ in pure 1T-TaS$_2$ is inconsistent with the previous results which report the absence of $\kappa_{0}/T$ \cite{TaS2_k}.   The vanishing of $\kappa_0/T$ in 1T-TaS$_{1.93}$Se$_{0.07}$ indicates that this discrepancy may be due to the difference of the degree of disorder between the present crystals and the crystals used in Ref.\cite{TaS2_k}. The present results suggest that the crystals with a small amount of disorder may be required to observe finite $\kappa_0/T$.

%Wiedemann-Franz law ensures that electron contribution in $\kappa_{0}/T$ is estimated less than 10$^{-5}$ mJmol$^{-1}$K$^{-2}$ and almost all part of finite temperature linear term is $\kappa_{spin}$, i.e. the presence of gapless spin excitations which are deconfined. 
%Such itinerant excitations in the QSLs have been attributed to emergent fractionalized quasiparticle excitations which carries spin but no charge. Moreover, the gapless excitations represented by finite $\kappa_{0}/T$ are consistent with a spinon Fermi surface, ruling out a Dirac spinon with nodes.  Our results for finite $\kappa_{0}/T$ does not support the previous study, in which $\kappa_{0}/T$ is absent. 
%For the thermal conductivity kappa of the doped and irradiated samples, it looks like the fit to kappa/T should be not quite linear. Indeed, locally-glassy physics would predict kappa/T ~ T^alpha with alpha near 1, but modified from it by logarithmic corrections that usually appear as a shift of 0.1--0.3 in the power law exponent. Usually in structural glasses at least, with phonon-dominated thermal conductivity, the logarithmic corrections suppress kappa so alpha is less than one. But it may be possible to get logarithmic enhancement in other settings.This will give further evidence that both doping and irradiation are captured well by models of local disorder/glassiness.

\subsection{Heat capacity of 1T-TaS$_{2-x}$Se$_{x}$}

\begin{figure}
\includegraphics[width=\columnwidth]{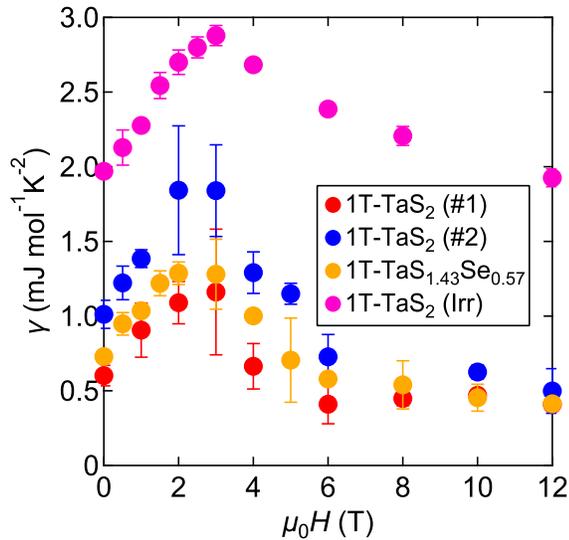}
\caption{\textcolor{black}{Magnetic field dependences of $\gamma$ for 1T-TaS$_2$\#1 (red), 1T-TaS$_2$\#2 (blue), 1T-TaS$_{1.43}$Se$_{0.57}$ (orrange) and 1T-TaS$_2$(Irr) (pink) obtained by the extrapolation above 1\,K. For all the systems, $\gamma$ increases with $H$ at low fields,  peaks and decreases at high fields.}}
\end{figure}

\begin{figure*}[t]
	\includegraphics[width=2\columnwidth]{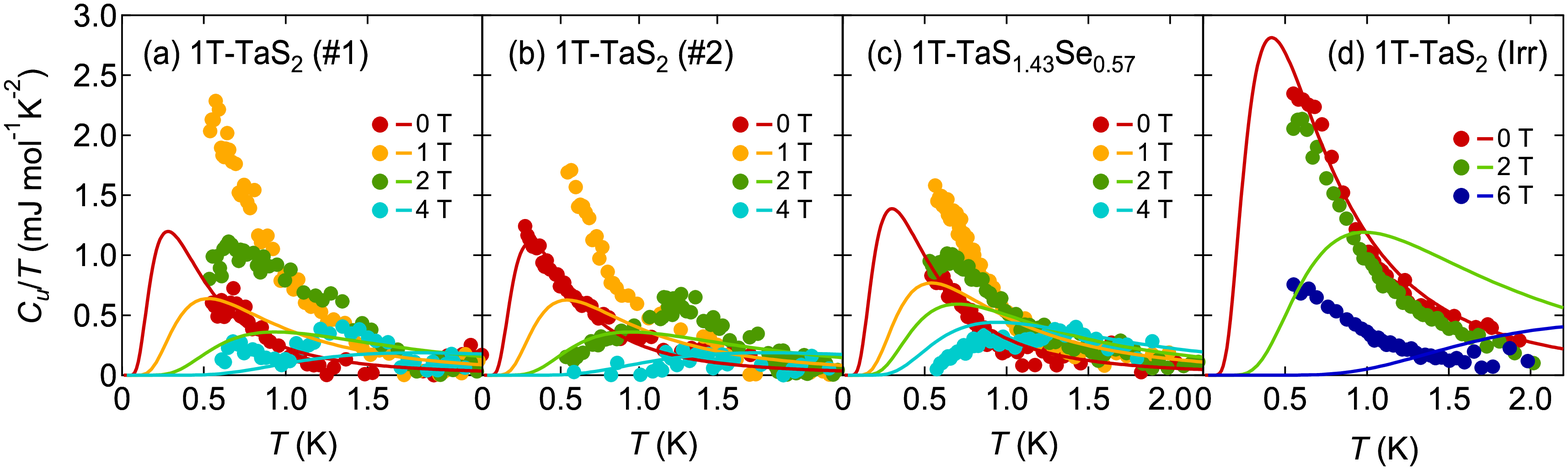}
	\caption{\textcolor{black}{Temperature dependence of $C_u/T=C/T-\gamma-\beta T^2$. Solid lines represent Schottky contribution $C_S$ calculated from Eq.\,(1). The parameters $A$ and $H_0$ in Eq.\,(1), both of which are field independent, are obtained from a fit to $C_u/T$ in zero field.}}
\end{figure*}

Figures\,4 (a)-(d) depict the heat capacity divided by temperature $C/T$ plotted as a function of $T^2$ in zero and finite magnetic fields applied perpendicular to the 2D plane for 1T-TaS$_2$(\#1, \#2), 1T-TaS$_{1.43}$Se$_{0.57}$,  and 1T-TaS$_2$(Irr), respectively.    In zero field,  $C/T$ exhibits an upturn at low temperatures for all the samples.  The upturn is enhanced at low fields, but it is largely suppressed at high fields.  \textcolor{black}{In high temperature regime, where no upturn is observed, $C/T$ increases in proportion to $T^2$.  There are two important features that should be emphasized in the heat capacity.  The first  feature is that  a simple extrapolation of $C/T$  to $T\rightarrow$0 from high temperature region above 1\,K shows finite intercept at $T=0$, demonstrating the presence of finite $\gamma T$ term that represents the gapless fermionic excitations.  Thus $C/T$ is well fitted as $C/T=\gamma+\beta T^2$ in the temperature range where upturn of $C/T$ is not observed.   For 1T-TaS$_2$ (\#1 and \#2) and 1T-TaS$_{1.43}$Se$_{0.57}$,  the upturn of $C/T$ at low temperatures is not observed at $\mu_0H\geq 4$\,T and   $C/T=\gamma+\beta T^2$ holds in the whole temperature range.    It is obvious that as shown in Fig.\,4(a), $\gamma$ in zero field is largely enhanced at $\mu_0H=2$\,T, while it is largely suppressed at  $\mu_0H=12$\,T. }  
	
\textcolor{black}{In Fig.\,5, the detailed field dependence of $\gamma(H)$ obtained by the extrapolation above 1\,K is shown.   For all crystals,  $\gamma$ initially increases with $H$, and then peaks and decreases.   In stark contrast to $\kappa/T$ which is only slightly enhanced by the magnetic field,  $\gamma$ is remarkably suppressed by the field at $\mu_0H\geq 3$\,T. %$\gamma$ in zero field is remarkably suppressed at $\mu_0H=12$\,T except for 1T-TaS$_2$(Irr). 
}  The thermal conductivity is governed by the itinerant excitations, while the heat capacity contains both localized and itinerant excitations.    Therefore these highly unusual contrasting response to magnetic field firmly establishes the presence of two types of gapless excitations with itinerant and localized characters.    In 1T-TaS$_{1.43}$Se$_{0.57}$ and 1T-TaS$_2$(Irr), $\gamma$ remains finite, while $\kappa_0/T$ is absent. These results indicate that the itinerant gapless excitations are sensitive to the disorder, while the localized gapless excitations are robust against the disorder.   

\textcolor{black}{The contrasting response of the field dependence to adding disorder therefore also indicates
%The contrasting response also indicates 
that the $\gamma$-term mainly arise from the localized spin excitations. 
Therefore, the $\gamma$-term should arise from some gapless localized excitations of spin singlets.   However, as a magnetic field has no effect on the singlet energy but rather breaks a singlet and turns the pair of spins into a triplet, 
the resulting $\gamma$-term would then be expected to be suppressed by a magnetic field. 
%the $\gamma$-term is expected to be suppressed by a magnetic field.   
Therefore, the low field enhancement of $\gamma$-term cannot be explained by a simple gapless localized spin-singlet model.} 

\begin{figure}[b]
	\includegraphics[width=0.7\columnwidth]{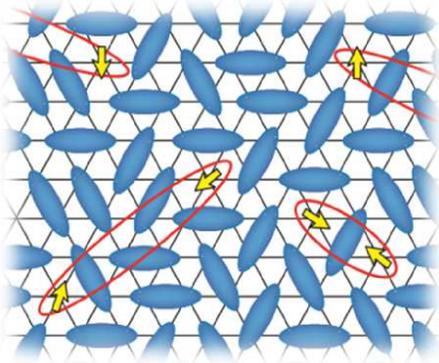}
	\caption{A schematic illustration of the proposed spin liquid state of 1T-TaS$_2$, which contains gapless excitations with localized and itinerant characters.  The localized excitations arise from orphan spins (yellow arrows) that form random long range valence bonds (red ellipsoids).  They are surrounded by \textcolor{black}{the proposed} spin liquid state (blue ellipsoids) with itinerant excitations,  which may be  attributed to  spinons that form a Fermi surface.}
\end{figure}

\begin{figure*}[t]
	\includegraphics[width=2\columnwidth]{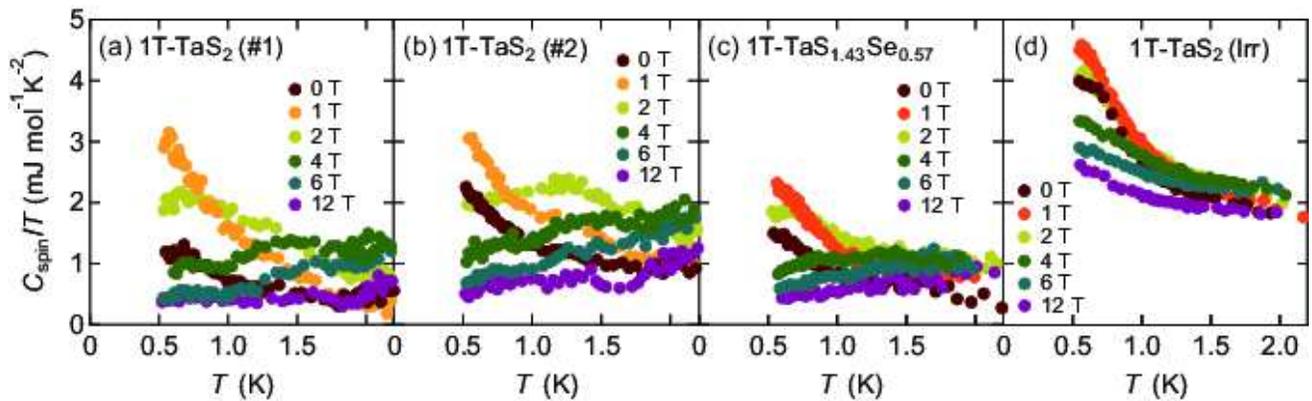}
	\caption{
	\textcolor{black}{Temperature dependence of $C_\mathrm{spin}/T$ for (a) 1T-TaS$_2$\#1, (b) 1T-TaS$_2$\#2, (c) 1T-TaS$_{1.43}$Se$_{0.57}$ and (d) 1T-TaS$_2$(Irr) in zero field and at $\mu_0H=$1, 2, 4, 6, ans 12\,T.} 	
		}
\end{figure*}

\begin{figure*}[t]
	\includegraphics[width=2\columnwidth]{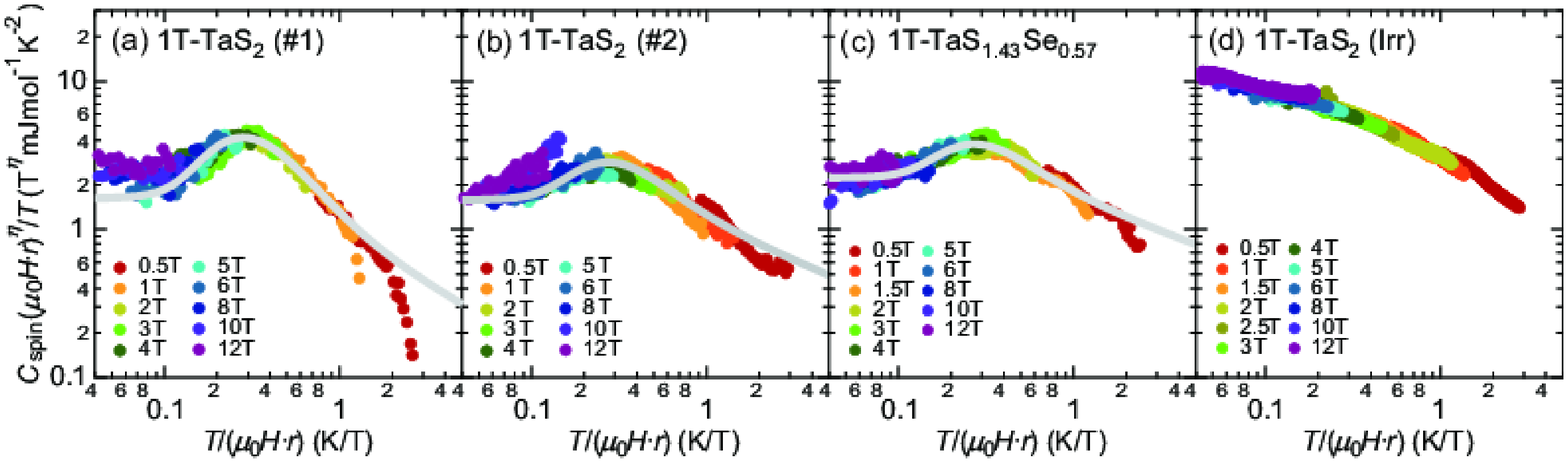}
	\caption{Scaling relationship of spin contribution of the heat capacity; $C_{spin}(\mu_0 H\cdot r)^{\eta}/T$ plotted as a function of $T/(\mu_0 H\cdot r)$ for (a) 1T-TaS$_2$\#1, (b) 1T-TaS$_2$\#2, (c) 1T-TaS$_{1.43}$Se$_{0.57}$ and (d) 1T-TaS$_2$(Irr).
Good scaling is observed with the parameters of $\eta=$ 0.68, $r=$1.37 for 1T-TaS$_2$\#1, $\eta=$ 0.44, $r=$1.49 for 1T-TaS$_2$\#2 and $\eta=$ 0.49, $r=$1.22 for 1T-TaS$_{1.43}$Se$_{0.57}$.  
Gray lines for (a)-(c) indicate the scaling function obtained by the fit to Eq.\,(2). 
For 1T-TaS$_2$(Irr), although all the data collapse into a single curve, $C_{spin}(\mu_0 H\cdot r)^{\eta}/T$ monotonically decreases with $T/(\mu_0 H\cdot r)$ and cannot be described by Eqs.(2) and (3). 	}
\end{figure*}

\textcolor{black}{The second important feature is the  the upturn of $C/T$ at low temperatures. } There are two scenarios for this upturn.  One is the nuclear  Schottky contribution $C_S$  and the other is the contribution of orphan spins that form random long range valence bonds.    We first analyze the data in accordance with the two-level Schottky model, where the upturn is described by 
\begin{equation}
C_{S}=A\left[\frac{\Delta(H)}{k_{B}T}\right]^{2}e^{-\Delta(H)/k_{B}T}. 
\end{equation}
\textcolor{black}{Here, $A$ is a constant that is proportional to the number of two-level systems %the crystal volume 
and   $\Delta(H)=g\mu_B\sqrt{H_0^2+H^2}$ is the energy of the excited level, where $H_0$ is the magnetic field that is characterized by the crystal electric field, $g$ %(=2) 
is the electron $g$-factor  which is assumed as 2 here 
and $\mu_B$ is the Bohr magneton.  
Figures\,6(a)-(d) depict the upturn component of the heat capacity at low temperature $C_u/T$, which is yielded by subtracting $\gamma$ and $\beta T^2$ from total heat capacity $C_u/T=C/T-\gamma-\beta T^2$.  Here  $\gamma$ and $\beta T^2$  are obtained from $C/T$ above \,1K, where the heat capacity is not influenced by the low temperature upturn.  %The fitting parameters are $A$ and $H_0$, both of which are field independent. 
We fit $C_u/T$ by Eq.\,(1) in zero field to determine $A$ and $H_0$, both of which are field independent, and then calculate $C_S$ in magnetic fields.   
As shown in Figs.6(a)-(d), %if we fit $C_u/T$ by Eq.\,(1) in zero field, 
$C_u/T$ in magnetic fields strongly deviates from the calculated $C_S$ (solid lines) for all the crystals.  %calculation for all crystals.  
We find that in order to fit $C_u/T$ in the whole field range, we need to assume strongly field dependent $A$, which is unphysical.  Thus the low temperature upturn of $C/T$ is unlikely to be due to the Schottky contribution. } 

%In Fig.\,5, we also plot $H$-dependence of $\gamma$ obtained by assuming that the low temperature enhancement of $C/T$ is due to Schttoky contribution and 

 %We find that, with parameters $\beta,\gamma$ considered separately at each magnetic field value,  $C/T$ as a function of temperature is well fitted by $C/T=\gamma+\beta T^2+ C_S/T$ in the whole temperature range for all the systems. 
%The fits use  $\Delta(0)/k_B=1.1, 0.6, 1.1, 1.5$\,K for 1T-TaS$_2$\#1, \#2, 1T-TaS$_{1.43}$Se$_{0.57}$, 1T-TaS$_2$(Irr), respectively. 
%\sout{In particular as shown in Figs.\,5 (a)-(d), $(C-C_S)/T$ is well fitted by the expression $\gamma +\beta T^2$.}  
%\textcolor{red}{Figures\,5(a)-(d) show the temperature dependence of the heat capacity after subtracting the Schottky contribution.} 
%Figure\,6 depicts the field dependence of $\gamma$.  For all systems, $\gamma$ increases with $H$ at low fields, peaks and decreases at high fields.

%\begin{figure}[t]
%	\includegraphics[width=0.7\columnwidth]{Fig6.eps}
%	\caption{Magnetic field dependences of $\gamma$ for 1T-TaS$_2$\#1 (red), 1T-TaS$_2$\#2 (blue), 1T-TaS$_{1.43}$Se$_{0.57}$ (orrange) and 1T-TaS$_2$(Irr) (purple) estimated by extrapolating  $(C - C_S)/T$  to $T = 0$, assuming that the low temperature enhancement of $C/T$ arises from the Schottky contribution. For all the systems, $\gamma$ increases with $H$ at low fields,  peaks and decreases at high fields.}
%\end{figure}

%The field dependence of $\gamma$ shown in Fig.\,6 is obtained by assuming that the low temperature enhancement of $C/T$ arises from the Schottky contribution.

 \textcolor{black}{Thus, to account for the unusual field dependence of the $\gamma$-term and low temperature enhancement of $C/T$, a new idea is required.} Recently, through new theoretical studies of the role of quenched disorder in quantum paramagnetic states, an alternative scenario has been proposed \cite{VBS_defect, QSL_defect_C}. 
In this picture,  as shown schematically in Fig.\,7, the majority of spin-1/2 sites form a quantum paramagnetic state such as a spin liquid, while a small fraction of sites host nucleated ``orphan spins", which are not microscopic defects but rather are emergent quantum objects that arise from a competition of disorder and frustration. %and are different spinons in a uniform QSL state.    
An orphan spin can couple with another orphan spin to form a singlet state.  The exchange energies between these orphan spins vary randomly, depending on their distance, which leads to a formation of singlets with random energy gaps whose distribution is exponentially broad. This broad distribution includes singlets with arbitrarily small energy gaps (red ellipsoids in Fig.\,7), leading to a state with gapless excitations.

These gapless excitations give rise to $\gamma$-term-like contributions in the heat capacity but similarly also contributes non-$T$-linear terms due to the non-trivial  energy distribution of the random singlets.  Combined, these contributions give rise to an enhancement of $C/T$ at low temperatures, in a similar way as the Schottky contribution but with a distinct field dependence.  In particular this model predicts that magnetic contribution $C_m$ in the heat capacity arising from the localized spin excitations collapse into a single curve of form,
\begin{equation}
\frac{C_m(H,T)}{T}\sim \frac{1}{H^{\eta}}F_q(T/H),
\end{equation} 
where $F_q(X)$ is a scaling function which is determined by  the energy distribution of the random singlets;  
\begin{eqnarray}
F_q(X) \sim \left\{
\begin{array}{ll}
X^q  &X \ll 1  \\
X^{-\eta}(1+c_0/X^2) &X \gg 1.
 \end{array}\right.
 \end{eqnarray}
Here $\eta$ is a non-universal exponent, $0\leq \eta \leq1$, that characterizes the random-singlet distribution, $q$ is determined by the spin and crystal symmetries with $q=0$ for 1T-TaS$_2$,  and $c_0$ is a constant.   When $\eta\neq 0$,  $F_0(X)$  increases with $X$, peaks   and  decreases at large $X$.   

Here we analyze the heat capacity in accordance with the random long-range valence bonds scenario.   To examine the scaling relation given by Eq.(2),  the magnetic part is obtained by subtracting the acoustic phonon contribution, $C_{spin}=C-\beta T^3$. \textcolor{black}{Here we use $\beta$ determined in zero field. Figures\,8(a)-(d) depicts the temperature dependence of $C_{spin}/T$. For 1T-TaS$_2$ (\#1, \#2) and 1T-TaS$_{1.43}$Se$_{0.57}$, $C_{spin}/T$ increases with decreasing temperature at $\mu_0H\geq2$\,T and is almost $T$-independent at the lowest temperature at $\mu_0H\geq4$\,T. On the other hand, in 1T-TaS$_2$(Irr), $C_{spin}/T$ shows an upturn in the whole field range.} 

In Figs.\,\textcolor{black}{9}(a)-(d),  $C_{spin}(\mu_0 H\cdot r)^{\eta}/T$ 
%$C_{spin}(H,T)H^{-\eta}$ 
is plotted as a function of $T/H$ for 1T-TaS$_2$\#1, \#2, 1T-TaS$_{1.43}$Se$_{0.57}$ and 1T-TaS$_2$(Irr), respectively, where $r=g/g_{0}$ with $g_0=2$ and the effective $g$-factor of 1T-TaS$_2$, $g$.   The fitting parameters are $\eta$ and $r$.   
We find that %all the 
\textcolor{black}{most of} data collapse into a single curve with $\eta=$ 0.68 and $r=$1.37 for 1T-TaS$_2$\#1, $\eta=$ 0.44 and $r=$1.49 for 1T-TaS$_2$\#2 and $\eta=$ 0.49 and $r=$1.22 for 1T-TaS$_{1.43}$Se$_{0.57}$.  
\textcolor{black}{Deviation from the scaling function, which is observed at high fields ($H\geq$10\,T) and at the lowest field measured (0.5\,T), can be ascribed to interplay of the local moments with the proposed spin liquid \cite{QSL_defect_C}, and also to an error in estimating $\beta$ ($\sim\pm0.1$\,mJ/molK$^4$). 
%can be ascribed at least in the high-field and low-temperature limit to interplay of the magnetized spins with the proposed spin liquid \cite{QSL_defect_C}, and also in both limits could be ascribed to an error in estimating $\beta$ ($\sim\pm0.1$\,mJ/molK$^4$)
%Deviation from the scaling function, which is observed at high fields ($e.g.,$ 10 and 12\,T) and 0.5 T, is ascribed to a relatively large error in estimating $\beta$ ($\sim\pm0.1$\,mJ/molK$^2$).
}  
The scaling function obtained by Eq.(2) is shown by gray lines.  Good scaling relation with using similar values of $\eta$ and $r$ suggest that three systems are in a similar QSL state.  These results, along with the presence of finite $\kappa_0/T$,  suggest the microscopic coexistence of  localized orphan spins that form random singlets and itinerant spinons that form Fermi surface.   We note that the scaling function Eq.(2) assume only the localized excitations.  Therefore the scaling collapse shown in Figs.\,\textcolor{black}{9}(a)-(c) implies that the gapless excitations observed in the heat capacity are dominated by localized ones, i.e. $C_{spin}\approx C_m$.  
Small contribution of the itinerant gapless quasiparticles to the heat capacity 
%\sout{would imply the small Fermi surface of spinon} 
%\textcolor{red}{
would reflect a large exchange energy scale in 1T-TaS$_2$ \cite{He_spinon}. %}. 
%distribution function is different among smples
%Small contribution of the itinerant gapless quasiparticles to the heat capacity would %appears to %imply the small Fermi surface of spinon. %distribution function is different among smples

Let us then estimate the mean free path of the itinerant spinons.  The thermal conductivity is written as $\kappa/T=\frac{1}{3}\gamma_sv_s\ell_s$, where $\gamma_s$ is the heat capacity coefficient,  $v_s$ is the velocity and $\ell_s$ are the mean free path of the itinerant quasiparticles.   Since gapless localized excitations dominate over gapless itinerant ones,  $\gamma_s\ll \gamma$,  the lower limit of $\ell_s$ is obtained.   Using $v_{s}\sim\frac{J}{h}$, where $J\sim 1500$\,K is the exchange energy of spins %\textcolor{red}{
\cite{TaS2_muon_NQR}, %}, 
and $\gamma\sim 0.4\,$mJ/K$^2$mol at $\mu_0H=$12\,T (this corresponds to $\sim$5.2\,mJ/K$^2$ per spin), we obtain $\ell_s\gg 5$\,nm, which is much longer than the inter-spin distance (1\,nm).   This long mean free path suggests that the itinerant spinons may be little scattered by the localized orphan spins,  which deserves future study. This long mean free path is also consistent with the result that $\kappa_{0}/T$ is sensitive to Se-substitution.

As shown in Fig.\,\textcolor{black}{9}(d), although all data collapse into a single curve in 1T-TaS$_2$(Irr), the scaling function monotonically decreases with $T/H$, which is essentially different from that of 1T-TaS$_2$ (\#1 and \#2) and 1T-TaS$_{1.43}$Se$_{0.57}$.  This scaling function cannot be described by Eqs.\,(2) and (3), as $F(X)$ increases with $X$ at low $X$.  These results suggest that the strong disorder introduced in the magnetic Ta-layers leads to a possible new quantum paramagnetic state with a new scaling function.  
%size of gamma

\section{IV. CONCLUSION}
To reveal the nature of the QSL state in 1T-TaS$_{2}$ with the 2D perfect triangular lattice, we investigate the effect of randomness due to quenched disorder through the systematic measurements of heat capacity and thermal conductivity in pure, Se-substituted and electron irradiated crystals.  
The finite residual linear term of the thermal conductivity $\kappa_0/T$ indicates the presence of gapless itinerant quasiparticle excitations, which is consistent with the fractionalized spinons that form a Fermi surface.   The conduction of this gapless itinerant  quasiparticles is strongly suppressed by  weak disorder.  The finite  temperature linear heat capacity coefficient $\gamma$ is also observed.  This $\gamma$  is robust against the disorder and  exhibits very different field dependence than $\kappa_0/T$, demonstrating the coexistence  of gapless excitations with itinerant and localized character.   We find the universal scaling collapse of magnetic contribution of the heat capacity, which provides  support that localized  orphan spins form random singlets.    These results appear to capture an essential feature of the QSL state of 1T-TaS$_2$;  localized orphan spins induced by disorder form random valence bonds and are surrounded by a QSL phase with the spinon Fermi surface.  This feature is preserved by the  introduction of weak quenched disorder by Se-substitution.  
The introduction of strong disorder by electron irradiation, on the other hand,  dramatically changes the spin liquid state, leading to a possible new quantum paramagnetic state. 
These unique features of 1T-TaS$_2$ provide new insights into the influence of quenched disorder on a QSL and 
demonstrate the ability to introduce distinct types of disorder with qualitatively different effects on QSL states.

\section*{Acknowledgments}
We are very grateful to H. Kawamura, K. T. Law, P.A. Lee, E.-G. Moon, K. Totsuka and M. Udagawa for many useful comments and suggestions. This work was supported by Grants-in-Aid for Scientific Research (KAKENHI) (Nos. 25000003, 18H01177, 18H01180 and 18H05227, 18K13511) and on Innovative Areas Topological Material Science (No. JP15H05852), Quantum Liquid Crystals (No.\ JP19H05824), and 3D Active-Site Science (No. 26105004) from Japan Society for the Promotion of Science (JPSJ). I.K. was supported by a National Research Council Fellowship through the National Institute of Standards and Technology.

\end{document}